\documentclass[aps,prd,groupedaddress,onecolumn,showpacs,10pt]{revtex4-1}
\usepackage{graphicx,color}
\usepackage{bm}
\usepackage{amssymb,amsmath,amsfonts, mathrsfs}

\newcommand{\be}{\begin{equation}}
\newcommand{\ee}{\end{equation}}
\newcommand{\ba}{\begin{array}}
\newcommand{\ea}{\end{array}}
\newcommand{\bqa}{\begin{eqnarray}}
\newcommand{\eqa}{\end{eqnarray}}

\begin{document}


\title{Possible molecular states in $B^{(*)}B^{(*)}$  scatterings}


\author{Meng-Ting Yu}
\affiliation{School of Physics, Southeast University, Nanjing 211189,
P.~R.~China}
\author{Zhi-Yong Zhou}
\email[Corresponding author:]{zhouzhy@seu.edu.cn}
\affiliation{School of Physics, Southeast University, Nanjing 211189,
P.~R.~China}
\author{Dian-Yong Chen}
\email[]{chendy@seu.edu.cn}
\affiliation{School of Physics, Southeast University, Nanjing 211189,
P.~R.~China}
\author{Zhiguang Xiao}
\email[]{xiaozg@ustc.edu.cn}
\affiliation{Interdisciplinary Center for Theoretical Study, University of Science
and Technology of China, Hefei, Anhui 230026, China}


\date{\today}

\begin{abstract}
We present that, if unitarizing the $B^{(*)}B^{(*)}$ scattering amplitudes in the constituent interchange model, one can find two bound state poles for $(I_{tot},S_{tot})=(0,1)$ $BB^*$ and $B^*B^*$ system, which corresponds to two $I(J^P)=0(1^+)$ doubly bottomed molecular states. Furthermore, it is noticed that the virtual states in $(1,0)$ $BB$, $(1,1)$ $BB^*$, $(1,0)$ $B^*B^*$, and $(1,2)$ $B^*B^*$ systems  could produce enhancements of the module squares of the scattering $T$-matrix just above the related thresholds, which might correspond to $I(J^P)=1(0^+)$, $1(1^+)$, and $1(2^+)$ doubly bottomed molecular states, respectively.  The calculation may be helpful for searching for the doubly bottomed molecular state in future experiments.

\end{abstract}


\maketitle
\section{Introduction}
Since the $X(3872)$ was observed by Belle in 2003~\cite{Choi:2003ue},
searches for exotic multiquark states beyond the conventional meson
classifications have attracted intense attentions both from
experimental and theoretical sides. In these years, many
unconventional hidden-charm or hidden-beauty states  have been
observed, as reviewed in
refs.~\cite{Guo:2017jvc,Chen:2016qju,Lebed:2016hpi}. Some of these
states run out of the predictions of the quark potential model and
can not be described by the naive quark model. As a
result, new configurations such as hadron molecules, tetraquark
states, hybrid states, are utilized to understand these exotic
states. Many of these states are regarded as belonging to the valence
quark configuration $Q\bar Q q\bar q$~(where $Q$ denotes $c$ or $b$
quarks and $q$ denotes light quarks), which implies that QCD can form
hadron states  in an unconventional way~\cite{Weinberg:2013cfa}.

After the observations of so many hidden-charmed or hidden-bottomed
states, there arises a natural question whether there exist similar
doubly-charmed or doubly-bottomed molecular states with valence quark
configuration $Q\bar q Q\bar q$. Recently, LHCb reported the doubly
charmed baryon state $\Xi_{cc}^{++}$~\cite{PhysRevLett.119.112001},
which inspired Karliner and Rosner to predict a doubly bottomed
tetraquark state to be about 215 MeV below the $B^-\bar B^{0*}$
threshold soon later~\cite{KarlinerPhysRevLett.119.202001}, while
Eichten and Quig predicted it to be 121 MeV below the $B^-\bar B^{0*}$
threshold~\cite{EichtenPhysRevLett.119.202002}.  In fact, theoretical
explorations of such doubly-heavy meson states have been carried on by
several groups for a long time, as was reviewed in
ref.~\cite{Liu:2019zoy}.  In 1986, it was found that the $bb\bar u\bar
d$ state could be bounded in a non-relativistic potential
model~\cite{Zouzou1986}. In 1999, Barnes $et.\ al.$ found that the
$I=0$ $BB^*$ channel could be attractive and form a $J^P=1^+$ bound
state by solving the schrodinger equation of the $BB$, $B B^*$, and
$B^*B^*$ systems\cite{BarnesPhysRevC.60.045202}. After that, different
approaches to investigate the possibility of forming doubly bottomed
or doubly charmed meson states are studied through  tetraquark
models~\cite{VjiandePhysRevD.79.074010,EbertPhysRevD.76.114015,Ming_2008,DuPhysRevD.87.014003,Luo2017,yangPhysRevD.80.114023,Feng:2013kea,MehenPhysRevD.96.094028,Richard:2018yrm,Cai:2019orb},
meson exchange
models\cite{Wang:2018atz,LiuPhysRevD.89.074015,Xu:2017tsr,PhysRevD.88.114008},
which usually calculate the binding energies by solving the
Schr$\ddot{\mathrm{o}}$dinger Equation with the potentials obtained in such models.
There are also some other calculations based on the QCD sum
rules\cite{Wang:2017dtg,Agaev:2018khe} or using the Lattice
simulations\cite{UKQCDPhysRevD.60.054012,NPLPhysRevD.76.114503,BicudoPhysRevD.87.114511,BrownPhysRevD.86.114506,bicudoPhysRevD.92.014507,bicudoPhysRevD.93.034501}.
These calculations basically paid their attentions to whether the bound
states could be formed in the particular channels and the results  in
different calculations are not the same though most of them
claim that there is a bound state in the $I=0$ $BB^*$
channel with very high possibility. However, there could be cases
where the near threshold structure in the lineshape could be caused by
a virtial state or by a combined effect of the virtual state and the
threshold, when the virtual state is very close to the threshold.
In fact, there are models claiming that the famous $X(3872)$ could be
a virtual state just below the threshold\cite{}. However, the method of solving the
Schr\"odinger equation or using lattice calculations could only produce the
bound state and  would not give the virtual states. To study the
possibility of the near threshold virtual states, one needs to have a
nonperturbative scattering amplitude and annalytically continue it to
the complex plane and study the pole structure of the amplitude.

In this paper, we just try to study the binding problem of $BB$, $B
B^*$, $B^*B^*$ systems by investigating the existence of poles in the
unitary meson-meson scattering matrix. As is well known, a bound
state will appear as a pole of the partial-wave $S$ matrix below the threshold on the physical
Riemann sheet, while a virtual
state will appear as a pole below the threshold on the second Riemann
sheet. The unitary amplitude is obtained using the K-matrix
method by unitarization of a Born approximation of the amplitude. The valence quark interchange model by Barnes and Swanson is
adopted to provide the Born approximation of the meson-meson
scattering amplitude~\cite{Barnes:1991em,BarnesPhysRevC.60.045202}.
Then, the unitarization of partial-wave amplitudes are derived and the poles of unitarized
scattering amplitudes could be extracted by analytically continuing
the energy to the complex plane. All the S-wave partial wave
amplitudes of $BB$, $B B^*$, $B^*B^*$  scatterings are analyzed and
related bound-state or virtual-state poles are searched for.

The paper is organized as follows: The Barnes-Swanson model is
introduced and discussed in Section II. The K-matrix unitarization
method for the partial wave Barnes-Swanson amplitude is derived in
Section III. Numerical results and discussions are devoted in Section
IV.

\section{The model}
In the constituent  interchange model developed by Barnes and Swanson~\cite{Barnes:1991em,Barnes:1999hs}, the meson-meson scattering amplitude is calculated by the (anti)quark-(anti)quark interactions by assuming the one-gluon-exchange~(OGE) color Coulomb interaction, spin-spin interaction, and linear scalar confinement interaction.
In the coordinate space, the effective interaction Hamiltonian is
\bqa
H_I=\sum_{ij}[(\frac{\alpha_s}{r_{ij}}-\frac{8\pi\alpha_s}{3m_i m_j}\vec S_i\cdot\vec S_j\delta(\vec r_{ij})-\frac{3b}{4}r_{ij})\sum_a\mathcal{F}^a(i)\cdot\mathcal{F}^a(j)],
\eqa
where $i,j$ represent valence quark or anti-quark in different initial
hadrons. The color generator $\mathcal{F}^a=\lambda^a/2$ for quarks
and $\mathcal{F}^a=-\lambda^{aT}/2$ for  anti-quarks.
Actually, to make the scheme consistent, the quark-quark interactions
have also been  used in determining the meson spectroscopy and their
wave functions, so the model has a  small free parameter space.

It is more convenient to construct the scattering amplitude in the
momentum space. First, the meson state is defined by the mock state to represent its wave function as
\begin{align*}
|A(n, { }^{2S+1}L_{J,M})(\vec
P)\rangle=&\sum_{M_L,M_S}\langle LM_LSM_S
|JM\rangle\int \mathrm{d}^3p\,
\psi_{nLM_L}(\vec p)\,\chi^{12}_{SM_S}\,\phi^{12}\,\omega^{12}
\\&\times\Big|q_1\Big(\frac
{m_1}{m_1+m_2}\vec P+\vec p\Big)\bar q_2\Big(\frac {m_2}{m_1+m_2}\vec
P-\vec p\Big)\Big\rangle,
\end{align*}
 $\chi^{12}$, $\phi^{12}$ and $\omega^{12}$  are the spin wavefunction, flavor wave
function and the color wave function, respectively. $p_1$ ($p_2$) and
$m_1$ ($m_2$)
are the momentum and mass of the quark (anti-quark).   $\vec P=\vec
p_1+\vec p_2$ is the momentum of the  center of mass,
and $\vec p=\frac{m_2\vec p_1-m_1\vec p_2}{m_1+m_2}$ is the
relative momentum. $\psi_{nLM_L}$ is the wave function for the meson,
$n$ being the radial quantum number. The normalization is $\langle
\vec{P}',\lambda'|\vec{P},\lambda\rangle=\delta_{\lambda\lambda'}\delta(\vec
P'-\vec P)$, where $\lambda$ represents the quantum numbers such as
$nJM$ and particle species.

\begin{figure}[t]%
\begin{center}%
\includegraphics[height=20mm]{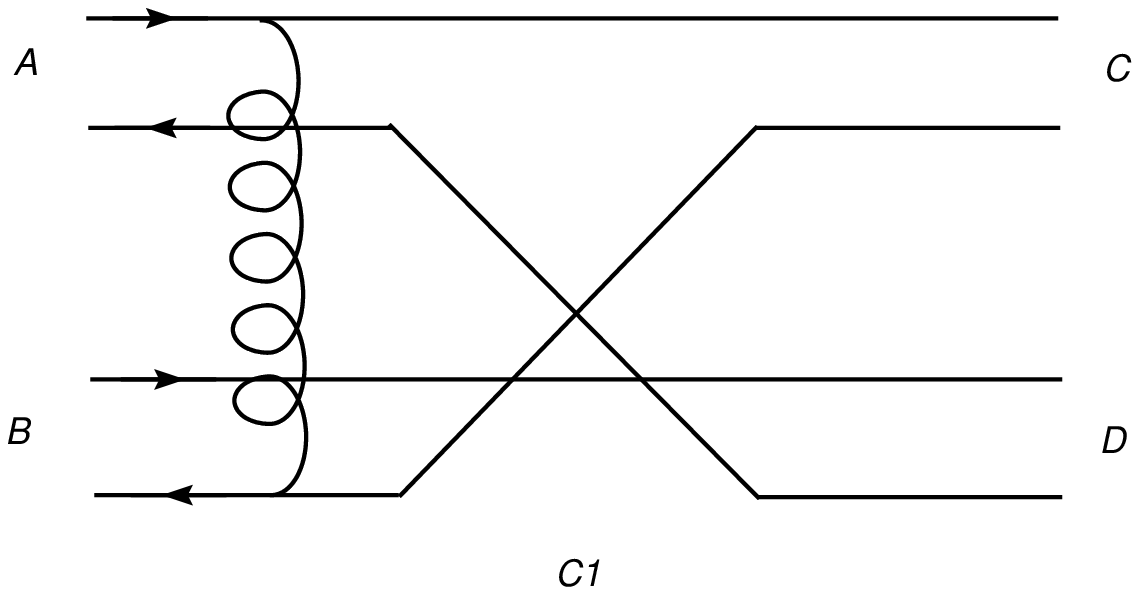}
\includegraphics[height=20mm]{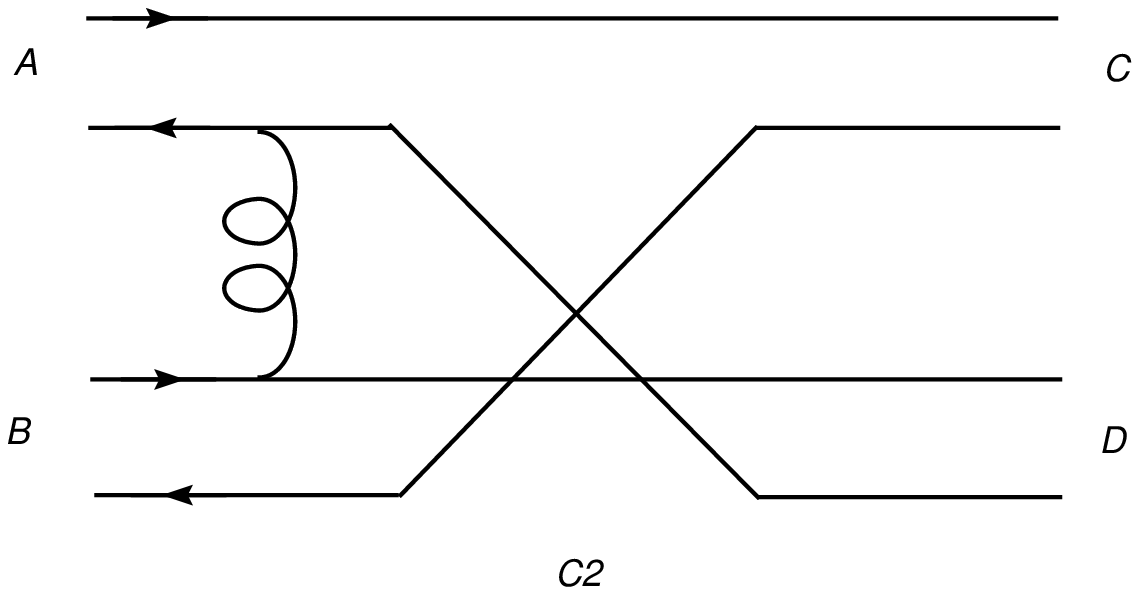}
\includegraphics[height=20mm]{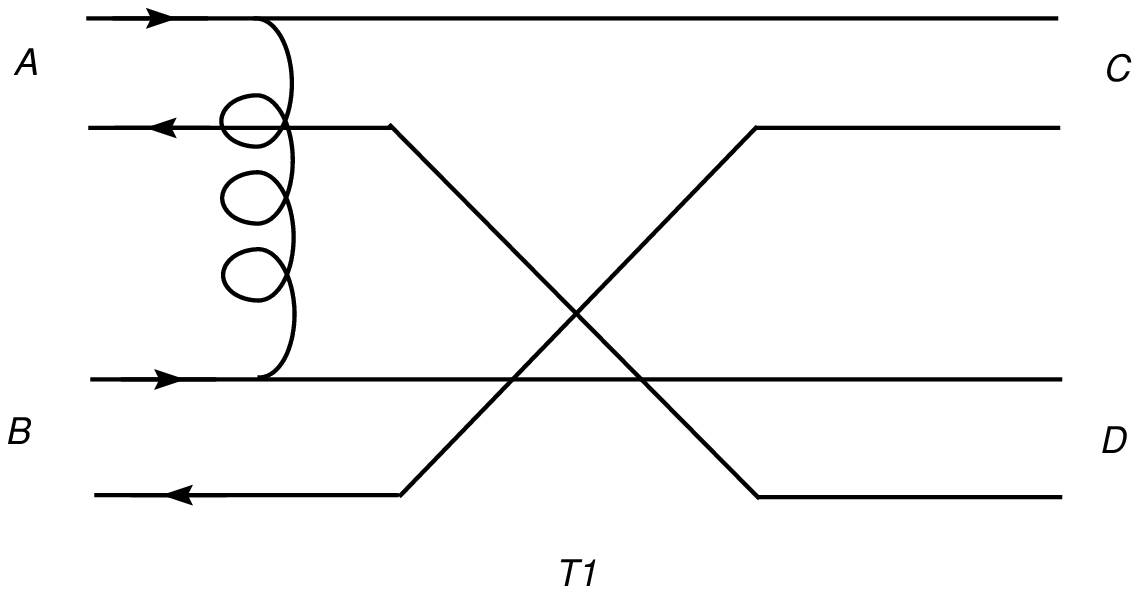}
\includegraphics[height=20mm]{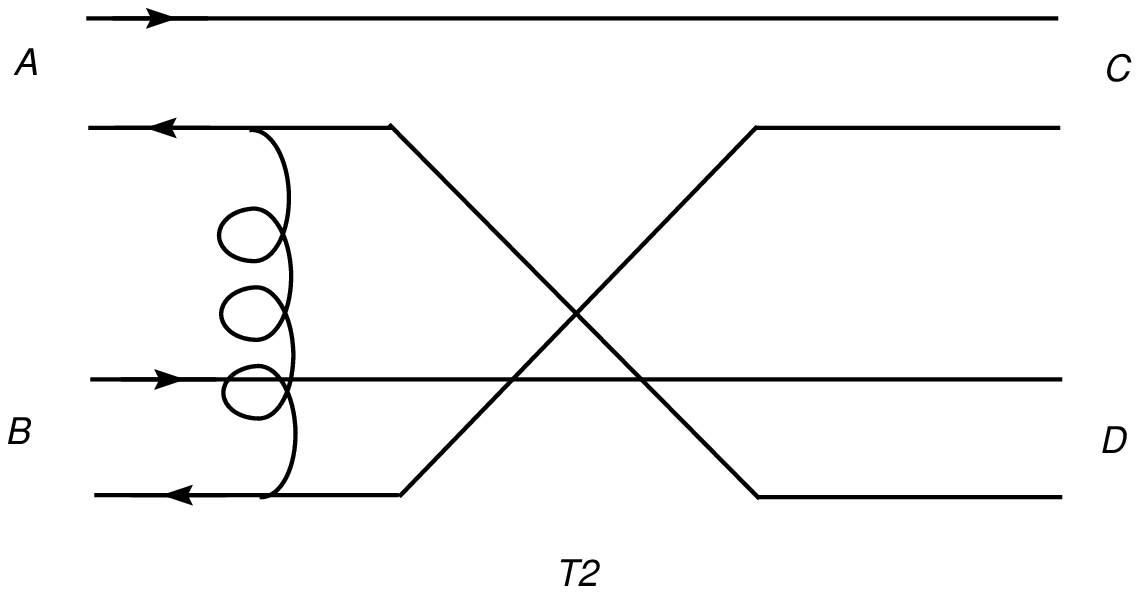}
\caption{\label{rearrange} The four quark rearrangement diagrams of $AB\rightarrow CD$ meson-meson scatterings for anti-quark exchange. The arrows represent the quark line directions.}
\end{center}%
\end{figure}

Then, at the lowest order the scattering amplitude for the process
$AB\rightarrow CD$ is just the matrix element of interaction $H_I$
between  the initial state $|A,B\rangle$ and  the final
$\langle C,D|$ which is formed just by rearrangements of the
quarks(anti-quarks) in the initial states. Thus, the $S$ matrix can be
written down as
\bqa
S_{fi}=\delta_{fi}-2\pi i \delta(E_f-E_i)\langle C,D|H_I|A,B\rangle
\eqa
where $i, f$ denotes the initial and final states, and the matrix element is expressed as
\bqa
\langle C,D|H_I|A,B\rangle=\delta^3(\vec C+\vec D-\vec A-\vec B)\mathcal{M}_{fi}
\eqa
where the $\vec C$ denotes the momentum of particle $C$, similarly for others.
Since the Hamiltonian is the interaction between the constituent
(anti-)quarks, the matrix element would be the integration of the
product of the spatial wave
functions of the (anti-)quarks in the mesons and the constituent
(anti-)quark scattering
amplitude.
To the Born order of the (anti-)quark scattering amplitude, there are
four kinds of $q\bar q-q\bar q$ scattering diagrams which are labeled
according to which pairs of the constituents are interacted. These are
 $q$-$\bar q$ interaction diagrams, i.e. ``capture$_1$"~($C_1$) and
``capture$_2$"~($C_2$), and $q$-$q$~($\bar q$-$\bar q$) interaction
diagrams, i.e. ``transfer$_1$"~($T_1$), and ``transfer$_2$"~($T_2$), as shown in
Fig.~\ref{rearrange}.

\begin{figure}[h]%
\begin{center}%
\includegraphics[height=20mm]{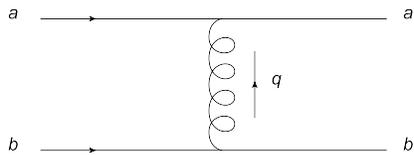}
\caption{\label{redefineq} Momentum redefinition in the quark-quark transition.}
\end{center}%
\end{figure}

To evaluate the contributions of these diagrams, it is more convenient
to redefine the momentum variables as in Fig.~\ref{redefineq}.
In the $q_a (\bar q_a) +q_b (\bar q_b)\to q_{a'} (\bar q_{a'})+ q_{b'} (\bar
q_{b'})$ quark(antiquark) transitions, the initial and final momenta
are denoted as $\vec a\vec b\to \vec a'\vec b'$. It is convenient to
define $\vec q=\vec a'-\vec a$, $\vec p=(\vec a'+\vec a)/2$.

Then, the  $T$-matrix element $\mathcal{M}_{fi}$ is contributed by the sum of all four kinds of diagrams, and the contribution of every diagram  could be written down as the product of signature, flavor, color, spin, and space factors, represented as  $h=I_{sign} I_{flavor} I_{color} I_{spin} I_{space}$. The
signature factor is $(-1)$ for all diagrams because of interchanging three quark lines as shown in Fig.\ref{rearrange}. The color factor is
$(-4/9)$
for two capture diagrams~($C_1$ and $C_2$) and $(4/9)$ for two transfer
diagrams~($T_1$ and $T_2$).
The spin factors for $C_1$, $C_2$, $T_1$, and $T_2$ diagrams in
different interaction terms are listed in
Table.~\ref{Table:spinfactor}.
\begin{table}[tp]
\begin{center}
\caption{\label{Table:spinfactor}Compilation of the spin factors for
$C_1$, $C_2$, $T_1$, and $T_2$ diagrams in the spin-spin hyperfine, color
Coulomb, and linear potential terms~\cite{Barnes:1991em}.}
\begin{tabular}{|c|c|c|c|c|c|c|c|c|c|c|c|c|}
\hline
$(S_A,S_B)\rightarrow (S_C,S_D)$ & \multicolumn{4}{|c|}{$(0,0)\rightarrow (0,0)$} & \multicolumn{4}{|c|}{{$(1,0)\rightarrow (1,0)$}}& \multicolumn{4}{|c|}{{$(0,1)\rightarrow (0,1)$}} \\
\hline
  & $C_1$ &$C_2$ & $T_1$&$T_2$& $C_1$ &$C_2$ & $T_1$&$T_2$ & $C_1$ &$C_2$ & $T_1$&$T_2$ \\
\hline
spin-spin & $-3/8$  & $ -3/8$  &  $3/8$    &  $3/8$ &   $1/8$ &   $-3/8$    &   $-1/8$    &     $3/8$  &   $-3/8$  &   $1/8$  &   $-1/8$    &     $3/8$   \\
\hline
Coulomb & \multicolumn{4}{|c|}{$1/2$  } & \multicolumn{4}{|c|}{{$1/2$}} & \multicolumn{4}{|c|}{{$1/2$}} \\
\hline
linear &\multicolumn{4}{|c|}{$1/2$  } & \multicolumn{4}{|c|}{{$1/2$}}& \multicolumn{4}{|c|}{{$1/2$}} \\
\hline
\hline
$(S_A,S_B)\rightarrow (S_C,S_D)$ & \multicolumn{4}{|c|}{$(1,1)\rightarrow (1,1),\ S_{tot}=2$} & \multicolumn{4}{|c|}{{$(1,1)\rightarrow (1,1),\ S_{tot}=1$}}& \multicolumn{4}{|c|}{{$(1,1)\rightarrow (1,1),\ S_{tot}=0$}} \\
\hline
  & $C_1$ &$C_2$ & $T_1$&$T_2$& $C_1$ &$C_2$ & $T_1$&$T_2$& $C_1$ &$C_2$ & $T_1$&$T_2$ \\
\hline
spin-spin &  $1/4$   & $ 1/4$  &  $1/4$    &  $1/4$  &   $0$ &   $0$    &   $-1/2$     &     $1/2$  &   $-1/8$  &   $-1/8$    &   $5/8$    &     $5/8$   \\
\hline
Coulomb & \multicolumn{4}{|c|}{$1$  } & \multicolumn{4}{|c|}{{$0$}} & \multicolumn{4}{|c|}{{$-1/2$}} \\
\hline
linear &\multicolumn{4}{|c|}{$1$  } & \multicolumn{4}{|c|}{{$0$}}& \multicolumn{4}{|c|}{{$-1/2$}} \\
\hline
\end{tabular}
\end{center}
\end{table}%
The space factor for each diagram is an overlap integral of the meson wave functions times the underlying  quark transition amplitude $T^{pot}_{fi}$. The overlap integrals of wave functions could be written down explicitly as
\bqa
I^{C_1}_{space}&&=(2\pi)^{-3}\int\mathrm{d}^3q\mathrm{d}^3p\Phi^*_C(\vec p+\frac{\vec q}{2}-\frac{1+\lambda}{2}\vec C)\Phi^*_D(\vec p-\frac{\vec q}{2}-\vec A-\frac{1-\lambda}{2}\vec C)\nonumber\\
&&\times T^{pot}_{fi}(\vec q)\Phi_A(\vec p-\frac{\vec q}{2}-\frac{1+\lambda}{2}\vec A)\Phi_B(\vec p-\frac{\vec q}{2}-\frac{1-\lambda}{2}\vec A-\vec C),\nonumber\\
I^{C_2}_{space}&&=(2\pi)^{-3}\int\mathrm{d}^3q\mathrm{d}^3p\Phi^*_C(-\vec p+\frac{\vec q}{2}+\vec A-\frac{1+\lambda}{2}\vec C)\Phi^*_D(-\vec p-\frac{\vec q}{2}-\frac{1-\lambda}{2}\vec C)\nonumber\\
&&\times T^{pot}_{fi}(\vec q)\Phi_A(-\vec p+\frac{\vec q}{2}+\frac{1-\lambda}{2}\vec A)\Phi_B(-\vec p+\frac{\vec q}{2}+\frac{1+\lambda}{2}\vec A-\vec C),\nonumber\\
I^{T_1}_{space}&&=(2\pi)^{-3}\int\mathrm{d}^3q\mathrm{d}^3p\Phi^*_C(\vec p+\frac{\vec q}{2}-\frac{1+\lambda}{2}\vec C)\Phi^*_D(\vec p-\frac{\vec q}{2}-\vec A-\frac{1-\lambda}{2}\vec C)\nonumber\\
&&\times T^{pot}_{fi}(\vec q)\Phi_A(\vec p-\frac{\vec q}{2}-\frac{1+\lambda}{2}\vec A)\Phi_B(\vec p+\frac{\vec q}{2}-\frac{1-\lambda}{2}\vec A-\vec C),\nonumber\\
I^{T_2}_{space}&&=(2\pi)^{-3}\int\mathrm{d}^3q\mathrm{d}^3p\Phi^*_C(-\vec p+\frac{\vec q}{2}+\vec A-\frac{1+\lambda}{2}\vec C)\Phi^*_D(-\vec p-\frac{\vec q}{2}-\frac{1-\lambda}{2}\vec C)\nonumber\\
&&\times T^{pot}_{fi}(\vec q)\Phi_A(-\vec p+\frac{\vec q}{2}+\frac{1-\lambda}{2}\vec A)\Phi_B(-\vec p-\frac{\vec q}{2}+\frac{1+\lambda}{2}\vec A-\vec C).\nonumber\\
\eqa
The contributions to the quark-quark amplitude $T^{pot}_{fi}$ by spin-spin, color Coulomb, and Linear confinement interactions read as
\bqa
T^{pot}_{fi}(\vec q)=\left\{\begin{array}{cc}
                -\frac{8\pi\alpha_s}{3m_1m_2}[\vec S_1\cdot\vec S_2] & \mathrm{Spin-spin}\\
                \frac{4\pi\alpha_s}{q^2}\mathbf{\textit{I}} & \mathrm{Coulomb} \\
                \frac{6\pi b}{q^4}\mathbf{\textit{I}} & \mathrm{Linear}
              \end{array}\right.
\eqa

We will deal with the scattering amplitudes of $B  B$, $B  B^*$, and
$B^* B^*$ systems.
For simplicity, the $u$ and $d$
quarks are assumed to have the same mass. To study the physically
allowed system  with a certain total isospin quantum number listed
in Table.\ref{Table:system}, the phase convention of the $B$ and $\bar
B$ meson isodoublets is chosen to be $\{|\bar{B}^0\rangle,|B^-\rangle\}=\{-|b\bar
d\rangle,|b\bar u\rangle\}$ and $\{|B^+\rangle,|B^0\rangle\}=\{-|u\bar
b\rangle,-|d\bar b\rangle\}$, and similar for the $B^*$ and $\bar B^*$
states. For isospin $I=0$ $BB$ system,
$|I=0,I_3=0\rangle=-\frac{1}{\sqrt{2}}(|b\bar d\rangle |b\bar
u\rangle-|b\bar u\rangle|b\bar d\rangle )$. For isospin $I=1$ $BB$
system, $|I=1,I_3=0\rangle=-\frac{1}{\sqrt{2}}(|b\bar d\rangle |b\bar
u\rangle+|b\bar u\rangle|b\bar d\rangle )$.

\begin{table}[htp]
\begin{center}
\caption{\label{Table:system} The physically allowed systems for $B^{(*)}B^{(*)}$.}
\begin{tabular}{ccccc}
\hline
\hline
system & Total isospin & \multicolumn{3}{c}{Total spin} \\
\hline
     &   $I_{tot}$ & $\ \ S_{tot}=0\ \ $& $\ \ S_{tot}=1\ \ $& $\ \ S_{tot}=2\ \ $\\
\hline
   $BB$  &   1 & even L &   &   \\
     &   0 & odd L &   &   \\
\hline
   $BB^*$  &   1 &   &  all L &   \\
     &   0 &   & all L  &   \\
\hline
   $B^*B^*$  &   1 & even L & odd L  &  even L \\
     &   0 & odd L &  even L & odd L  \\
\hline
\hline
\end{tabular}
\end{center}
\end{table}%
For $BB$ system,  the scattering amplitude with the total isospin $I=0$ could be expressed explicitly as
\bqa\label{eq:BBI0}
\langle BB|T|BB\rangle_{I=0}
&=&\frac{1}{2}(\langle b\bar d| \langle b\bar u|-\langle b\bar u|\langle b\bar d| )T(|b\bar d\rangle |b\bar u\rangle-|b\bar u\rangle|b\bar d\rangle )\nonumber\\
&=&
-\frac{1}{2}\langle b\bar d| \langle b\bar u|T|b\bar u\rangle|b\bar d\rangle
-\frac{1}{2}\langle b\bar u|\langle b\bar d| T|b\bar d\rangle |b\bar u\rangle
+\frac{1}{2}\langle b\bar d| \langle b\bar u|T|b\bar d\rangle |b\bar u\rangle
+\frac{1}{2}\langle b\bar u|\langle b\bar d| T|b\bar u\rangle|b\bar d\rangle
 \nonumber\\
&=&-T_{fi}(AB\to CD)+T_{fi}(AB\to DC).
\eqa
The latter two terms of the second line in Eq.(\ref{eq:BBI0}) are
related to the ``symmetrized diagrams" with the quark-line exchange
rather than the anti-quark line exchange~\cite{Barnes:1991em}, which
has the effects of interchanging the final $C$ and $D$ mesons.
Including both kinds of diagrams also keeps the amplitudes for $BB$ scattering to be Bose-symmetric.
Similarly, for $I=1$ $BB$ scattering, the amplitude could be also written down as
\bqa
\langle BB|T|BB\rangle_{I=1}=T_{fi}(AB\to CD)+T_{fi}(AB\to DC).
\eqa
 The $B^*B^*$ scatterings are similar to the $BB$ scatterings, so
there is no need to write the relations down explicitly.

For the $BB^*\to BB^*$ scatterings, when I=0, the scattering amplitude could be expressed as
\bqa
&\langle BB^*|T|BB^*\rangle_{I=0}=\frac{1}{2}[-\langle \bar B^0B^{-*}|T| B^-\bar B^{0*}\rangle-\langle  B^-\bar B^{0*}|T| \bar B^0B^{-*}\rangle+\langle \bar B^0B^{-*}|T| \bar B^0B^{-*}\rangle+\langle B^-\bar B^{0*}|T|  B^-\bar B^{0*}\rangle]
\nonumber\\
&=-\langle \bar B^0B^{-*}|T| B^-\bar B^{0*}\rangle+\langle \bar B^0B^{-*}|T| \bar B^0B^{-*}\rangle.
\eqa
When I=1, the scattering amplitude reads as
\bqa
&\langle BB^*|T|BB^*\rangle_{I=1}=\frac{1}{2}[\langle \bar B^0B^{-*}|T| \bar B^0B^{-*}\rangle+\langle B^-\bar B^{0*}|T|  B^-\bar B^{0*}\rangle+\langle \bar B^0B^{-*}|T| B^-\bar B^{0*}\rangle+\langle  B^-\bar B^{0*}|T| \bar B^0B^{-*}\rangle]
\nonumber\\
&=\langle \bar B^0B^{-*}|T| B^-\bar B^{0*}\rangle+\langle \bar B^0B^{-*}|T| \bar B^0B^{-*}\rangle.
\eqa
Since the incoming two mesons are not identical, there is no
Bose symmetry in this scattering amplitude. One have to notice that
the latter term $\langle \bar B^0B^{-*}|T| \bar B^0B^{-*}\rangle$,
which is contributed by the diagrams with quark line interchanged, is
not just the amplitude with the two final mesons interchanged in
the first term.

\section{Partial wave decomposition and unitarization}

In the previous section, we only calculated the lowest order
scattering amplitude, which contains different partial wave component
and does not satisfy the nonperturbative unitarity. Thus it needs
partial wave decomposition and unitarization. Here we  briefly
describe the partial-wave decomposition and unitarity relation in our
convention.

In
general, consider the scattering process of $12\to 1'2'$, where all
particle are massive. 
We use $\vec k$($k'$) to denote the momentum of partical $1$ ($1'$) in the
initial(final) states in the center of
mass system and $E_1,E_2$ ($E_1', E_2'$) to denote the energies for
the two initial (final) particles with the total energy $E=E_1+E_2$.
The spins and the third components are denoted as $s_1$, $\sigma_1$
for particle 1 and so on.
By using the convention of \cite{Weinberg:1995mt}, the amplitude could be expanded in partial waves as
\bqa
&\mathcal M_{\vec k'\sigma_1',-\vec k'\sigma_2',\vec k\sigma_1,-\vec k\sigma_2}=(|\vec k||\vec k'|E_1E_2E_1'E_2'/EE')^{-1/2}\sum_{j\sigma
l'm's'\mu'lms\mu}C_{s_1s_2}(s,\mu;\sigma_1,\sigma_2)C_{ls}(j,\sigma;m,\mu)
\nonumber\\
&C_{s_1's_2'}(s',\mu';\sigma_1',\sigma_2')C_{l's'}(j,\sigma;m',\mu')Y_{l'}^{m'}(\hat k')Y_{l}^{m*}(\hat k)M^j_{l's',ls}
\eqa
If we only considered the
spin-spin, color Coulomb, and linear confinement interactions here,
the total orbital angular momentum and the total spin are conserved separately.
The partial wave amplitude with total angular momentum $j$ will be
\bqa
&M^j_{ls,ls}=2\pi (|\vec k||\vec k'|E_1E_2E_1'E_2'/EE')^{1/2}\sum_{\sigma_1\sigma_2\sigma_1'\sigma_2'}C_{s_1s_2}(s,\mu;\sigma_1,\sigma_2)\nonumber\\ &\times C_{s_1's_2'}(s,\mu;\sigma_1',\sigma_2')
\int \mathrm{d} \mathrm{cos}\tilde\theta
P_l(\mathrm{cos}\tilde\theta)\mathcal M_{\vec k'\sigma_1',-\vec k'\sigma_2',\vec k\sigma_1,-\vec k\sigma_2}
\eqa
where $\tilde\theta$ is the angle between $\vec k$ and $\vec k'$. That means, in this calculation, the partial-wave elastic unitarity condition is as simple as that of the scalar particles
\bqa
\mathrm{Im}[ M^j_{ls}]=\pi M^{j*}_{ls}M^j_{ls},
\eqa
where we have omitted the repeated subscript $ls$ for brevity and $\mathrm{Im}[\ \ ]$ means the imaginary part of the related function.
If we redefine the amplitude $M^j_{ls}=\frac{1}{\pi}\rho t_l$ by extracting the kinematic factor $\rho=\frac{2|\vec k|}{E}$, one will obtain a familiar form similar to the elastic unitarity condition of partial waves for scalar particles as
\bqa
\mathrm{Im}[t_l]=\rho t_l^* t_l
\eqa
or in a more concise form
\bqa
\mathrm{Im}[t_l^{-1}] =-\rho.
\eqa
In the constituent interchange model, only the Born term is
calculated, so there is no elastic cut in the scattering amplitude and
it does not obey the unitarity relation. One need to restore it by
adopting a suitable unitarization scheme. Here we use the
 K-matrix unitarization method, which could be regarded as summing
over all the bubble chains. Then, the unitarized partial-wave $S$
matrix element could be represented as
\bqa
S_l=1+2i T=\frac{1+i\rho t_l }{1-i\rho t_l },
\eqa
and the scattering $T$-matrix element is
\bqa
T_l=\frac{\rho t_l }{1-i\rho t_l }.
\eqa
The pole of $S$-matrix element below the threshold on the real axis of
the first Riemann sheet, satisfying $1-i\rho t_l=0$, represents a
bound state. When the unitarity relation is satisfied, the scatteing $S$-matrix on the second Riemann sheet is the inverse of that on the first Riemann sheet, $S^{\mathrm{2nd\ sheet}}=1/S^{\mathrm{1st\ sheet}}$, that means, the zero point
of the first Riemann sheet corresponds to the pole of the second
sheet. Thus, the zero point satisfying $1+i\rho t_l=0$ below
the threshold represents a virtual state.

\section{Numerical calculations and discussions}
The parameters in the calculation is provided by the
Godfrey-Isgur~(GI) model~\cite{Godfrey:1985xj} because its
interactions are similar to those in the Barnes-Swanson model and it
presented a generally successful prediction to the meson mass
spectrum. In the GI model, the wave functions of mesons are expanded
in a series of a very large number of harmonics oscillator~(HO) wave
functions, which make it difficult to decompose the amplitude in the
angular momentum in an analytical form, so we approximate the meson
wave function by a HO wave function carrying the same radial quantum
number and orbital angular momentum as the meson, with its effective
radius obtained by the rms radius $r_{rms}$ of the related meson state
in the GI model.

The running coupling function is parameterized as
$\alpha_s(q^2)=0.25e^{-q^2}+0.15e^{-\frac{q^2}{10}}+0.20e^{-\frac{q^2}{1000}}$  to saturate the result of perturbation theory calculation in the large $q^2$ region and avoid the divergence in the low $q^2$ region,
and the quark masses used here are $m_u=m_d=0.22$GeV, $m_b=4.977$GeV.
The strength coefficient of the confinement linear potential is
$b=0.18$. We used the HO wave functions, with the oscillator
parameters of the bottomed mesons as $\beta_{B}=0.579$GeV,
$\beta_{B^*}=0.542$GeV~($\beta$ is defined in the HO wave function by $\psi(\vec r)\sim
(\mathrm{polynomial})\times e^{-\beta^2 r^2/2}$).

In this calculation, only the partial $S$-wave
 scatterings of $B^{(*)}B^{(*)}$  are investigated.  
The scattering systems are labeled by their total isospins and total spins
as $(I_{tot},S_{tot})$. Thus, the $B^{(*)}B^{(*)}$ systems discussed
here, which could have non-vanishing partial $S$-waves, are $(1,0)$
$BB$, $(1,1)$ $BB^*$, $(0,1)$ $BB^*$, $(1,0)$ $B^*B^*$, $(0,1)$
$B^*B^*$, $(1,2)$ $B^*B^*$, as listed in Table.~\ref{Table:system}.
Using the standard parameters listed above,
there are a bound state found in $(0,1)$ $BB^*$ system and $(0,1)$
$B^*B^*$ system respectively, and one virtual pole is found in each of
the other systems, whose pole positions are listed in Table.\ref{Table:poles}.
All the bound states and virtual states are just near the thresholds
of the related channels.

\begin{figure}[t]%
\caption{\label{Tmatrix} The absolute squares of scattering $T$-matrix of $(I_{tot},S_{tot})=(1,0)$ $BB$, $(1,1)$ $BB^*$, $(0,1)$ $BB^*$, $(1,0)$ $B^*B^*$, $(0,1)$ $B^*B^*$, and $(1,2)$ $B^*B^*$ systems.}
\begin{center}%
\includegraphics[height=30mm]{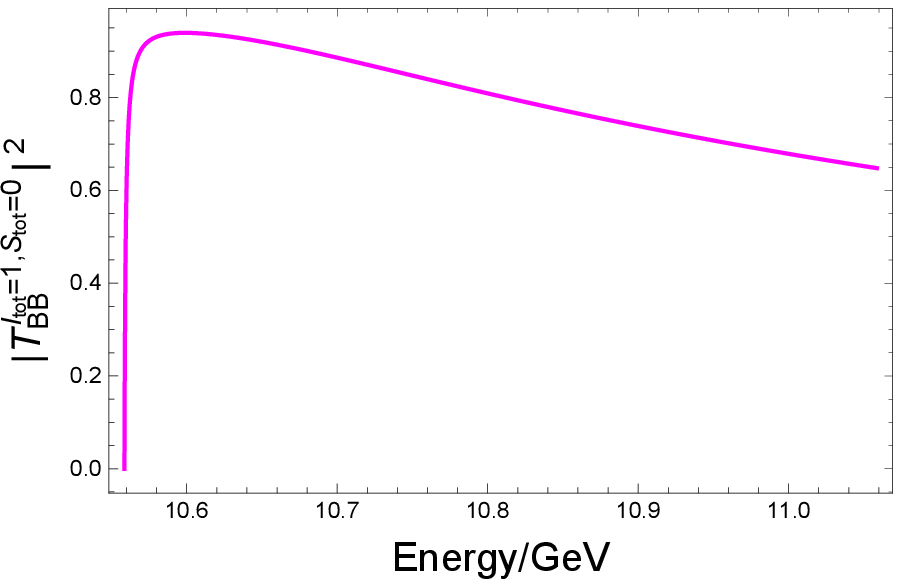}
\includegraphics[height=30mm]{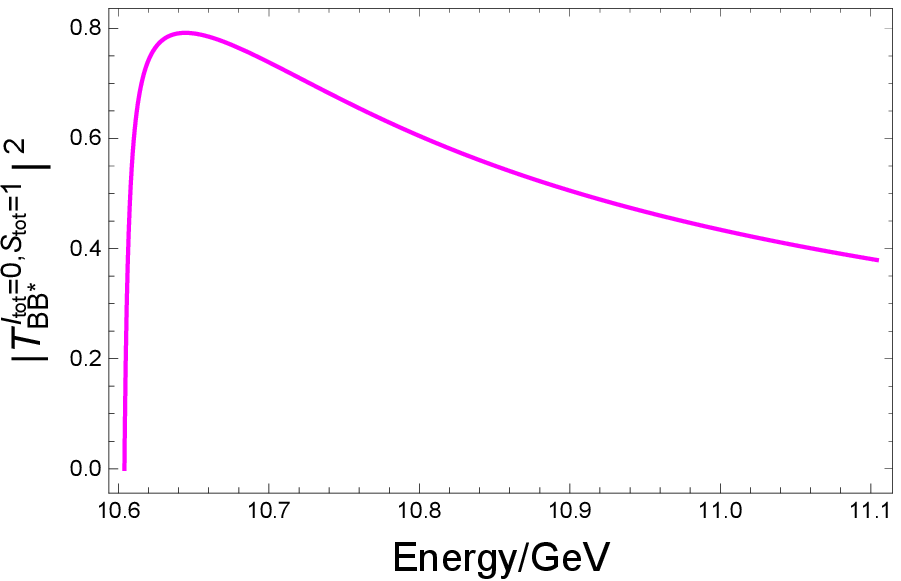}
\includegraphics[height=30mm]{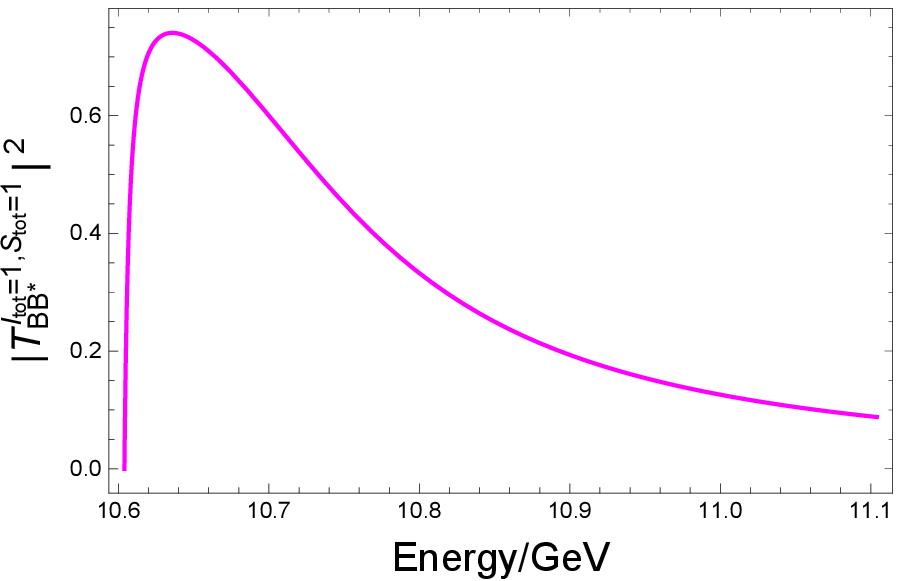}
\\
\includegraphics[height=30mm]{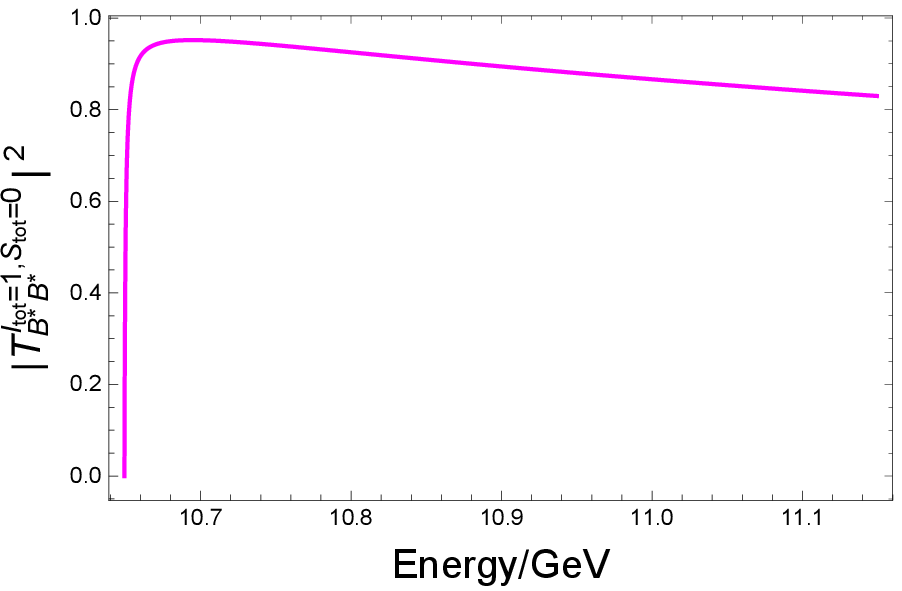}
\includegraphics[height=30mm]{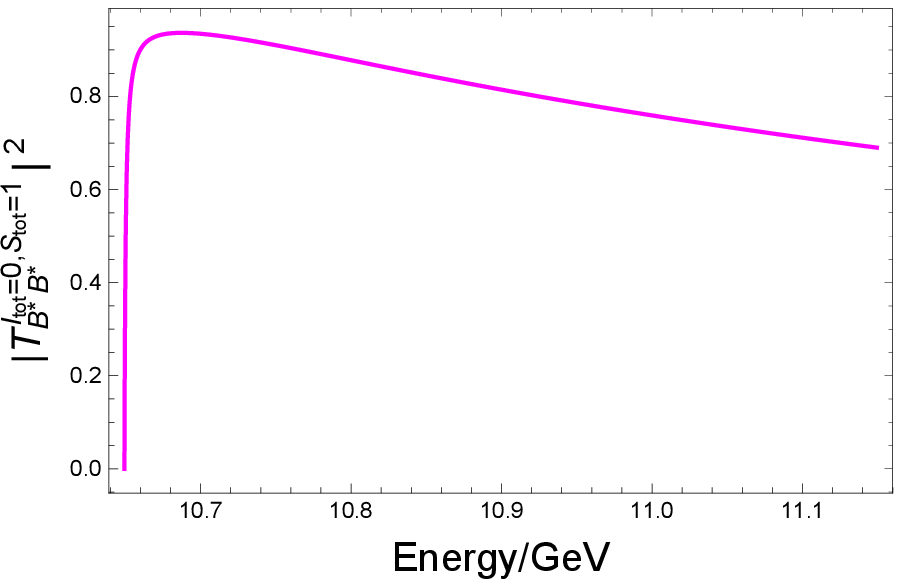}
\includegraphics[height=30mm]{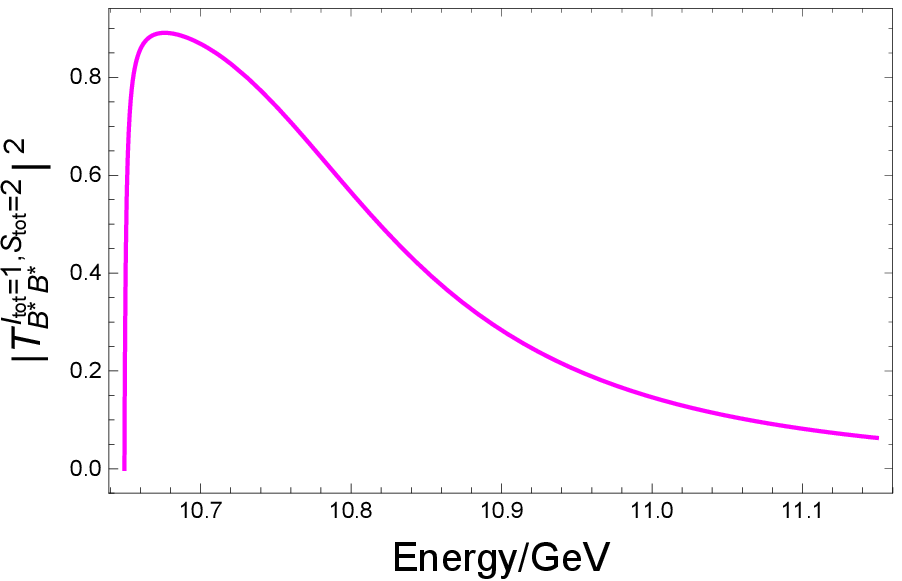}

\end{center}%
\end{figure}

Usually, if there is a virtual state close enough to the
threshold, it will produce an enhancement for the absolute square of the
scattering amplitude $|T|^2$ just above the related thresholds, which
could be discerned in experiments. Whether the threshold enhancement
could be recogonized as a state also depends on the interplay between
the wave functions and the threshold. As shown in Fig.\ref{Tmatrix},
although there is a near threshold virtual state in each of the
$(1,0)$ $BB$, $(1,1)$ $BB^*$, $(1,0)$ $B^*B^*$, and $(1,2)$ $B^*B^*$
systems, the near threshold peaks of $|T|^2$ in $(1,1)$ $BB^*$ and
$(1,2)$ $B^*B^*$ systems are more obvious than the other two even
though the
virtuals states in the latter are closer to the related thresholds
than the former.

In Ref.~\cite{Barnes:1999hs}, the authors extracted  approximated
local potentials from a part of the Born amplitudes which are calculated using the
BS model. Then by
solving the two-meson schrodinger equation with these potentials,
they found that only the $I=0$ $BB^*$ channel is
attractive enough to form a $I(J^P)=0(1^+)$ bound state with a binding energy of
$-5.5$ MeV, which is consistent with our result.
We also found that the $(0,1)$ $B^*B^*$ system could also form another $I(J^P)=0(1^+)$ bound state.
Since our
unitarization approach is using the full Born amplitude and the wave
functions and coupling function used in the calculation are also
different from the method In Ref.~\cite{Barnes:1999hs}, the
differences between the results may not be surprising.  In addition,
our method provide more informations about the appearance of the
virtual states.
Another approach using the heavy meson chiral effective theory in
Ref.~\cite{Wang:2018atz}
also found that the $I(J^P)=0(1^+)$ $BB^*$ and $B^*B^*$ systems are
attractive and obtain two bound states with their binding energies to
be $\Delta E_{BB^*}\sim -12.6^{+9.2}_{-12.9}$ MeV and $\Delta
E_{B^*B^*}\sim -23.8^{+16.3}_{-21.5}$ MeV,
respectively. This may indicate a strong interaction in both of the $(0,1)$
$BB^*$ and  the  $(0,1)$ $B^*B^*$ system, which is similar in our result.

\begin{table}[t]
\begin{center}
\caption{\label{Table:poles} The pole positions of the physically allowed systems for $B^{(*)}B^{(*)}$. The subscript ``$v$" denotes a virtual state and ``$b$"  a bound state.}
\begin{tabular}{ccccc}
\hline
\hline
system(threshold)  & Total isospin & \multicolumn{3}{c}{Total spin} \\
\hline
     &   $I_{tot}$ & $\ \ S_{tot}=0\ \ $& $\ \ S_{tot}=1\ \ $& $\ \ S_{tot}=2\ \ $\\
\hline
   $BB$  &   1 & $E_v=10557.8$MeV &   &   \\
     ($E_{th}=10558.6$MeV) &   0 &  &   &   \\
\hline
   $BB^*$  &   1 &   &  $E_v=10600.6$MeV  &   \\
 ($E_{th}=10604.0$MeV)    &   0 &   & $E_b=10600.9$MeV   &   \\
\hline
   $B^*B^*$  &   1 & $E_v=10648.7$MeV  &    &
$E_v=10648.3$MeV \\
($E_{th}=10649.4$MeV)     &   0 &   &  $E_b=10648.6$MeV  &    \\
\hline
\hline
\end{tabular}
\end{center}
\end{table}%

There is a simple qualitative argument for the appearance of the
virtual states in different channels in our calculation. We have
stated that the virtual state would produce a near threshold
enhancement in the $|T|^2$
to be observed in the experiments, which means that there is a large
scattering length in these processes. In fact, from the theoretical point
of view, it is the large scattering length calculated from BS model
that causes the existence of the virtual states. In the K-matrix
formalism, the virtual state pole position is the solution to $1+i\rho
t_l=0$. If the perturbative scattering amplitude $t_l$ at the
threshold, which is just proportional to the scattering length, is
positive and large enough, $1+i\rho t_l=0$ will be saturated near the
threshold, because the kinematic factor $i\rho$ is negative below
the threshold and approaches to zero at the threshold. We can also see
that the larger the scattering length is, the nearer the position of
solution is to the threshold.  Because of the large scattering length,
$|T|^2$ will usually present a peak just above the threshold, which is
observed in the experiments as a state. Similarly, if $t_l$ is
negative  and its absolute value is large, it will imply a
near-threshold bound state. But for the deep bound state, it will
depend on the detailed behavior of $t_l$ below the threshold. Thus, in
general, as long as a model can produce a scattering length with a large
enough absolute value, there would be a near-threshold virtual state
or bound state generated.
Thus, it is the large absolute values of the
scattering lengths calculated from the BS model in these systems, as shown in Table \ref{scatteringlength}, that
are essential for the presence of the near threshold virtual and bound
states. Since BS model is a nonrelativistic model, we would expect that the
near threshold property, in particular the largeness  of the
scattering length, is qualitatively correct. Thus the appearance of
the near threshold virtual states might be a more general result.

\begin{table}%
\begin{center}
\caption{\label{scatteringlength} The scattering lengthes of different $B^{(*)}B^{(*)}$ scatterings. Unit is $\mathrm{GeV}^{-1}$.}
\begin{tabular}{ccccc}
\hline
\hline
system(threshold)  & Total isospin & \multicolumn{3}{c}{Total spin} \\
\hline
     &   $I_{tot}$ & $\ \ S_{tot}=0\ \ $& $\ \ S_{tot}=1\ \ $& $\ \ S_{tot}=2\ \ $\\
\hline
   $BB$  &   1 & 14.8 &   &   \\
     ($E_{th}=10558.6$MeV) &   0 &  &   &   \\
\hline
   $BB^*$  &   1 &   &  7.01  &   \\
 ($E_{th}=10604.0$MeV)    &   0 &   & -7.39   &   \\
\hline
   $B^*B^*$  &   1 & 16.3  &    & 12.8 \\
($E_{th}=10649.4$MeV)     &   0 &   &  -15.0  &    \\
\hline
\hline
\end{tabular}
\end{center}
\end{table}%

A further remark about the two $I(J^P)=0(1^+)$ states
generated in  $BB^*$ and $B^*B^*$ systems is in order. Since only the $S$-wave
amplitudes are considered in our calculation and different systems are
decoupled, the two states in $(I_{tot}=0,S_{tot}=1)$ $BB^*$ and
$B^*B^*$ actually have the same quantum numbers but are different
states. If these two channels are coupled to each other in reality,
they could appear in both channels and may affect each other.

\section{Summary}
In this paper, we use the Barnes-Swanson constituent interchange model
to provide the Born term of $B^{(*)}B^{(*)}$ scattering amplitude and
then unitarize them using the K-matrix method. By analytically
continuing the unitarized scattering amplitudes to the complex energy
plane, the dynamically generated bound states or virtual states could
be extracted. Two bound states with $I(J^P)=0(1^+)$ are found at about
10600.9 MeV in $BB^*$ scattering and at about
10648.6 MeV in $B^*B^*$ scattering, respectively. We also found that several near threshold virtual states could be found in $(1,0)$ $BB$, $(1,1)$ $BB^*$, $(1,0)$ $B^*B^*$, and $(1,2)$ $B^*B^*$ systems, which might produce
threshold enhancements  and correspond to  doubly bottomed molecular states.

In comparison, the other methods used in discussing
the states in these systems, such as solving the schr\"odinger
equation~\cite{Barnes:1999hs} using the potentials extracted from the
amplitude, or the heavy meson chiral effective theory
discussion~\cite{Wang:2018atz},  could only
produce the near threshold bound state information and can not say anything about
virtual states.
Besides the bound states, our method also predicts the near-threshold
virtual states in these systems which may have observable effects and
could also inspire more interests in searching for threshold
enhancements in experimental explorations.  We also provide a simple
qualitative explanation that it is the large scattering length
calculated from the BS model that is essential in generating these
virtual states which may be a more general result. From these
different approaches, we expect that the existence of the near
threshold composite states may be a general result in the $S$-wave
$BB$, $BB^*$ and $B^*B^*$ systems.

\begin{acknowledgments}
Helpful discussions with Yan-Rui Liu are appreciated.
This work is supported by China National Natural
Science Foundation under contract No. 11975075, No.  11105138, No. 11575177 and No.11235010. It is also supported by  the Natural Science Foundation of Jiangsu Province of China under contract No. BK20171349.
\end{acknowledgments}

\bibliographystyle{apsrev4-1}
\bibliography{Ref}

\begin{thebibliography}{37}%
\makeatletter
\providecommand \@ifxundefined [1]{%
 \@ifx{#1\undefined}
}%
\providecommand \@ifnum [1]{%
 \ifnum #1\expandafter \@firstoftwo
 \else \expandafter \@secondoftwo
 \fi
}%
\providecommand \@ifx [1]{%
 \ifx #1\expandafter \@firstoftwo
 \else \expandafter \@secondoftwo
 \fi
}%
\providecommand \natexlab [1]{#1}%
\providecommand \enquote  [1]{``#1''}%
\providecommand \bibnamefont  [1]{#1}%
\providecommand \bibfnamefont [1]{#1}%
\providecommand \citenamefont [1]{#1}%
\providecommand \href@noop [0]{\@secondoftwo}%
\providecommand \href [0]{\begingroup \@sanitize@url \@href}%
\providecommand \@href[1]{\@@startlink{#1}\@@href}%
\providecommand \@@href[1]{\endgroup#1\@@endlink}%
\providecommand \@sanitize@url [0]{\catcode `\\12\catcode `\$12\catcode
  `\&12\catcode `\#12\catcode `\^12\catcode `\_12\catcode `\%12\relax}%
\providecommand \@@startlink[1]{}%
\providecommand \@@endlink[0]{}%
\providecommand \url  [0]{\begingroup\@sanitize@url \@url }%
\providecommand \@url [1]{\endgroup\@href {#1}{\urlprefix }}%
\providecommand \urlprefix  [0]{URL }%
\providecommand \Eprint [0]{\href }%
\providecommand \doibase [0]{http://dx.doi.org/}%
\providecommand \selectlanguage [0]{\@gobble}%
\providecommand \bibinfo  [0]{\@secondoftwo}%
\providecommand \bibfield  [0]{\@secondoftwo}%
\providecommand \translation [1]{[#1]}%
\providecommand \BibitemOpen [0]{}%
\providecommand \bibitemStop [0]{}%
\providecommand \bibitemNoStop [0]{.\EOS\space}%
\providecommand \EOS [0]{\spacefactor3000\relax}%
\providecommand \BibitemShut  [1]{\csname bibitem#1\endcsname}%
\let\auto@bib@innerbib\@empty
\bibitem [{\citenamefont {Choi}\ \emph {et~al.}(2003)\citenamefont {Choi} \emph
  {et~al.}}]{Choi:2003ue}%
  \BibitemOpen
  \bibfield  {author} {\bibinfo {author} {\bibfnamefont {S.~K.}\ \bibnamefont
  {Choi}} \emph {et~al.} (\bibinfo {collaboration} {Belle Collaboration}),\
  }\href {\doibase 10.1103/PhysRevLett.91.262001} {\bibfield  {journal}
  {\bibinfo  {journal} {Phys. Rev. Lett.}\ }\textbf {\bibinfo {volume} {91}},\
  \bibinfo {pages} {262001} (\bibinfo {year} {2003})},\ \Eprint
  {http://arxiv.org/abs/hep-ex/0309032} {arXiv:hep-ex/0309032 [hep-ex]}
  \BibitemShut {NoStop}%
\bibitem [{\citenamefont {Guo}\ \emph {et~al.}(2018)\citenamefont {Guo},
  \citenamefont {Hanhart}, \citenamefont {Mei{\ss}ner}, \citenamefont {Wang},
  \citenamefont {Zhao},\ and\ \citenamefont {Zou}}]{Guo:2017jvc}%
  \BibitemOpen
  \bibfield  {author} {\bibinfo {author} {\bibfnamefont {F.-K.}\ \bibnamefont
  {Guo}}, \bibinfo {author} {\bibfnamefont {C.}~\bibnamefont {Hanhart}},
  \bibinfo {author} {\bibfnamefont {U.-G.}\ \bibnamefont {Mei{\ss}ner}},
  \bibinfo {author} {\bibfnamefont {Q.}~\bibnamefont {Wang}}, \bibinfo {author}
  {\bibfnamefont {Q.}~\bibnamefont {Zhao}}, \ and\ \bibinfo {author}
  {\bibfnamefont {B.-S.}\ \bibnamefont {Zou}},\ }\href {\doibase
  10.1103/RevModPhys.90.015004} {\bibfield  {journal} {\bibinfo  {journal}
  {Rev. Mod. Phys.}\ }\textbf {\bibinfo {volume} {90}},\ \bibinfo {pages}
  {015004} (\bibinfo {year} {2018})},\ \Eprint
  {http://arxiv.org/abs/1705.00141} {arXiv:1705.00141 [hep-ph]} \BibitemShut
  {NoStop}%
\bibitem [{\citenamefont {Chen}\ \emph {et~al.}(2016)\citenamefont {Chen},
  \citenamefont {Chen}, \citenamefont {Liu},\ and\ \citenamefont
  {Zhu}}]{Chen:2016qju}%
  \BibitemOpen
  \bibfield  {author} {\bibinfo {author} {\bibfnamefont {H.-X.}\ \bibnamefont
  {Chen}}, \bibinfo {author} {\bibfnamefont {W.}~\bibnamefont {Chen}}, \bibinfo
  {author} {\bibfnamefont {X.}~\bibnamefont {Liu}}, \ and\ \bibinfo {author}
  {\bibfnamefont {S.-L.}\ \bibnamefont {Zhu}},\ }\href {\doibase
  10.1016/j.physrep.2016.05.004} {\bibfield  {journal} {\bibinfo  {journal}
  {Phys. Rept.}\ }\textbf {\bibinfo {volume} {639}},\ \bibinfo {pages} {1}
  (\bibinfo {year} {2016})},\ \Eprint {http://arxiv.org/abs/1601.02092}
  {arXiv:1601.02092 [hep-ph]} \BibitemShut {NoStop}%
\bibitem [{\citenamefont {Lebed}\ \emph {et~al.}(2017)\citenamefont {Lebed},
  \citenamefont {Mitchell},\ and\ \citenamefont {Swanson}}]{Lebed:2016hpi}%
  \BibitemOpen
  \bibfield  {author} {\bibinfo {author} {\bibfnamefont {R.~F.}\ \bibnamefont
  {Lebed}}, \bibinfo {author} {\bibfnamefont {R.~E.}\ \bibnamefont {Mitchell}},
  \ and\ \bibinfo {author} {\bibfnamefont {E.~S.}\ \bibnamefont {Swanson}},\
  }\href {\doibase 10.1016/j.ppnp.2016.11.003} {\bibfield  {journal} {\bibinfo
  {journal} {Prog. Part. Nucl. Phys.}\ }\textbf {\bibinfo {volume} {93}},\
  \bibinfo {pages} {143} (\bibinfo {year} {2017})},\ \Eprint
  {http://arxiv.org/abs/1610.04528} {arXiv:1610.04528 [hep-ph]} \BibitemShut
  {NoStop}%
\bibitem [{\citenamefont {Weinberg}(2013)}]{Weinberg:2013cfa}%
  \BibitemOpen
  \bibfield  {author} {\bibinfo {author} {\bibfnamefont {S.}~\bibnamefont
  {Weinberg}},\ }\href {\doibase 10.1103/PhysRevLett.110.261601} {\bibfield
  {journal} {\bibinfo  {journal} {Phys. Rev. Lett.}\ }\textbf {\bibinfo
  {volume} {110}},\ \bibinfo {pages} {261601} (\bibinfo {year} {2013})},\
  \Eprint {http://arxiv.org/abs/1303.0342} {arXiv:1303.0342 [hep-ph]}
  \BibitemShut {NoStop}%
\bibitem [{\citenamefont {Aaij}\ \emph {et~al.}(2017)\citenamefont {Aaij},
  \citenamefont {Adeva}, \citenamefont {Adinolfi}, \citenamefont {Ajaltouni},
  \citenamefont {Akar},\ and\ \citenamefont
  {Albrecht}}]{PhysRevLett.119.112001}%
  \BibitemOpen
  \bibfield  {author} {\bibinfo {author} {\bibfnamefont {R.}~\bibnamefont
  {Aaij}}, \bibinfo {author} {\bibfnamefont {B.}~\bibnamefont {Adeva}},
  \bibinfo {author} {\bibfnamefont {M.}~\bibnamefont {Adinolfi}}, \bibinfo
  {author} {\bibfnamefont {Z.}~\bibnamefont {Ajaltouni}}, \bibinfo {author}
  {\bibfnamefont {S.}~\bibnamefont {Akar}}, \ and\ \bibinfo {author}
  {\bibnamefont {Albrecht}} (\bibinfo {collaboration} {LHCb Collaboration}),\
  }\href {\doibase 10.1103/PhysRevLett.119.112001} {\bibfield  {journal}
  {\bibinfo  {journal} {Phys. Rev. Lett.}\ }\textbf {\bibinfo {volume} {119}},\
  \bibinfo {pages} {112001} (\bibinfo {year} {2017})}\BibitemShut {NoStop}%
\bibitem [{\citenamefont {Karliner}\ and\ \citenamefont
  {Rosner}(2017)}]{KarlinerPhysRevLett.119.202001}%
  \BibitemOpen
  \bibfield  {author} {\bibinfo {author} {\bibfnamefont {M.}~\bibnamefont
  {Karliner}}\ and\ \bibinfo {author} {\bibfnamefont {J.~L.}\ \bibnamefont
  {Rosner}},\ }\href {\doibase 10.1103/PhysRevLett.119.202001} {\bibfield
  {journal} {\bibinfo  {journal} {Phys. Rev. Lett.}\ }\textbf {\bibinfo
  {volume} {119}},\ \bibinfo {pages} {202001} (\bibinfo {year}
  {2017})}\BibitemShut {NoStop}%
\bibitem [{\citenamefont {Eichten}\ and\ \citenamefont
  {Quigg}(2017)}]{EichtenPhysRevLett.119.202002}%
  \BibitemOpen
  \bibfield  {author} {\bibinfo {author} {\bibfnamefont {E.~J.}\ \bibnamefont
  {Eichten}}\ and\ \bibinfo {author} {\bibfnamefont {C.}~\bibnamefont
  {Quigg}},\ }\href {\doibase 10.1103/PhysRevLett.119.202002} {\bibfield
  {journal} {\bibinfo  {journal} {Phys. Rev. Lett.}\ }\textbf {\bibinfo
  {volume} {119}},\ \bibinfo {pages} {202002} (\bibinfo {year}
  {2017})}\BibitemShut {NoStop}%
\bibitem [{\citenamefont {Liu}\ \emph {et~al.}(2019)\citenamefont {Liu},
  \citenamefont {Chen}, \citenamefont {Chen}, \citenamefont {Liu},\ and\
  \citenamefont {Zhu}}]{Liu:2019zoy}%
  \BibitemOpen
  \bibfield  {author} {\bibinfo {author} {\bibfnamefont {Y.-R.}\ \bibnamefont
  {Liu}}, \bibinfo {author} {\bibfnamefont {H.-X.}\ \bibnamefont {Chen}},
  \bibinfo {author} {\bibfnamefont {W.}~\bibnamefont {Chen}}, \bibinfo {author}
  {\bibfnamefont {X.}~\bibnamefont {Liu}}, \ and\ \bibinfo {author}
  {\bibfnamefont {S.-L.}\ \bibnamefont {Zhu}},\ }\href {\doibase
  10.1016/j.ppnp.2019.04.003} {\bibfield  {journal} {\bibinfo  {journal} {Prog.
  Part. Nucl. Phys.}\ }\textbf {\bibinfo {volume} {107}},\ \bibinfo {pages}
  {237} (\bibinfo {year} {2019})},\ \Eprint {http://arxiv.org/abs/1903.11976}
  {arXiv:1903.11976 [hep-ph]} \BibitemShut {NoStop}%
\bibitem [{\citenamefont {Zouzou}\ \emph {et~al.}(1986)\citenamefont {Zouzou},
  \citenamefont {Silvestre-Brac}, \citenamefont {Gignoux},\ and\ \citenamefont
  {Richard}}]{Zouzou1986}%
  \BibitemOpen
  \bibfield  {author} {\bibinfo {author} {\bibfnamefont {S.}~\bibnamefont
  {Zouzou}}, \bibinfo {author} {\bibfnamefont {B.}~\bibnamefont
  {Silvestre-Brac}}, \bibinfo {author} {\bibfnamefont {C.}~\bibnamefont
  {Gignoux}}, \ and\ \bibinfo {author} {\bibfnamefont {J.~M.}\ \bibnamefont
  {Richard}},\ }\href {\doibase 10.1007/BF01557611} {\bibfield  {journal}
  {\bibinfo  {journal} {Zeitschrift f{\"u}r Physik C Particles and Fields}\
  }\textbf {\bibinfo {volume} {30}},\ \bibinfo {pages} {457} (\bibinfo {year}
  {1986})}\BibitemShut {NoStop}%
\bibitem [{\citenamefont {Barnes}\ \emph
  {et~al.}(1999{\natexlab{a}})\citenamefont {Barnes}, \citenamefont {Black},
  \citenamefont {Dean},\ and\ \citenamefont
  {Swanson}}]{BarnesPhysRevC.60.045202}%
  \BibitemOpen
  \bibfield  {author} {\bibinfo {author} {\bibfnamefont {T.}~\bibnamefont
  {Barnes}}, \bibinfo {author} {\bibfnamefont {N.}~\bibnamefont {Black}},
  \bibinfo {author} {\bibfnamefont {D.~J.}\ \bibnamefont {Dean}}, \ and\
  \bibinfo {author} {\bibfnamefont {E.~S.}\ \bibnamefont {Swanson}},\ }\href
  {\doibase 10.1103/PhysRevC.60.045202} {\bibfield  {journal} {\bibinfo
  {journal} {Phys. Rev. C}\ }\textbf {\bibinfo {volume} {60}},\ \bibinfo
  {pages} {045202} (\bibinfo {year} {1999}{\natexlab{a}})}\BibitemShut
  {NoStop}%
\bibitem [{\citenamefont {Vijande}\ \emph {et~al.}(2009)\citenamefont
  {Vijande}, \citenamefont {Valcarce},\ and\ \citenamefont
  {Barnea}}]{VjiandePhysRevD.79.074010}%
  \BibitemOpen
  \bibfield  {author} {\bibinfo {author} {\bibfnamefont {J.}~\bibnamefont
  {Vijande}}, \bibinfo {author} {\bibfnamefont {A.}~\bibnamefont {Valcarce}}, \
  and\ \bibinfo {author} {\bibfnamefont {N.}~\bibnamefont {Barnea}},\ }\href
  {\doibase 10.1103/PhysRevD.79.074010} {\bibfield  {journal} {\bibinfo
  {journal} {Phys. Rev. D}\ }\textbf {\bibinfo {volume} {79}},\ \bibinfo
  {pages} {074010} (\bibinfo {year} {2009})}\BibitemShut {NoStop}%
\bibitem [{\citenamefont {Ebert}\ \emph {et~al.}(2007)\citenamefont {Ebert},
  \citenamefont {Faustov}, \citenamefont {Galkin},\ and\ \citenamefont
  {Lucha}}]{EbertPhysRevD.76.114015}%
  \BibitemOpen
  \bibfield  {author} {\bibinfo {author} {\bibfnamefont {D.}~\bibnamefont
  {Ebert}}, \bibinfo {author} {\bibfnamefont {R.~N.}\ \bibnamefont {Faustov}},
  \bibinfo {author} {\bibfnamefont {V.~O.}\ \bibnamefont {Galkin}}, \ and\
  \bibinfo {author} {\bibfnamefont {W.}~\bibnamefont {Lucha}},\ }\href
  {\doibase 10.1103/PhysRevD.76.114015} {\bibfield  {journal} {\bibinfo
  {journal} {Phys. Rev. D}\ }\textbf {\bibinfo {volume} {76}},\ \bibinfo
  {pages} {114015} (\bibinfo {year} {2007})}\BibitemShut {NoStop}%
\bibitem [{\citenamefont {Ming}\ \emph {et~al.}(2008)\citenamefont {Ming},
  \citenamefont {Hai-Xia},\ and\ \citenamefont {Zong-Ye}}]{Ming_2008}%
  \BibitemOpen
  \bibfield  {author} {\bibinfo {author} {\bibfnamefont {Z.}~\bibnamefont
  {Ming}}, \bibinfo {author} {\bibfnamefont {Z.}~\bibnamefont {Hai-Xia}}, \
  and\ \bibinfo {author} {\bibfnamefont {Z.}~\bibnamefont {Zong-Ye}},\ }\href
  {\doibase 10.1088/0253-6102/50/2/31} {\bibfield  {journal} {\bibinfo
  {journal} {Communications in Theoretical Physics}\ }\textbf {\bibinfo
  {volume} {50}},\ \bibinfo {pages} {437} (\bibinfo {year} {2008})}\BibitemShut
  {NoStop}%
\bibitem [{\citenamefont {Du}\ \emph {et~al.}(2013)\citenamefont {Du},
  \citenamefont {Chen}, \citenamefont {Chen},\ and\ \citenamefont
  {Zhu}}]{DuPhysRevD.87.014003}%
  \BibitemOpen
  \bibfield  {author} {\bibinfo {author} {\bibfnamefont {M.-L.}\ \bibnamefont
  {Du}}, \bibinfo {author} {\bibfnamefont {W.}~\bibnamefont {Chen}}, \bibinfo
  {author} {\bibfnamefont {X.-L.}\ \bibnamefont {Chen}}, \ and\ \bibinfo
  {author} {\bibfnamefont {S.-L.}\ \bibnamefont {Zhu}},\ }\href {\doibase
  10.1103/PhysRevD.87.014003} {\bibfield  {journal} {\bibinfo  {journal} {Phys.
  Rev. D}\ }\textbf {\bibinfo {volume} {87}},\ \bibinfo {pages} {014003}
  (\bibinfo {year} {2013})}\BibitemShut {NoStop}%
\bibitem [{\citenamefont {Luo}\ \emph {et~al.}(2017)\citenamefont {Luo},
  \citenamefont {Chen}, \citenamefont {Liu}, \citenamefont {Liu},\ and\
  \citenamefont {Zhu}}]{Luo2017}%
  \BibitemOpen
  \bibfield  {author} {\bibinfo {author} {\bibfnamefont {S.-Q.}\ \bibnamefont
  {Luo}}, \bibinfo {author} {\bibfnamefont {K.}~\bibnamefont {Chen}}, \bibinfo
  {author} {\bibfnamefont {X.}~\bibnamefont {Liu}}, \bibinfo {author}
  {\bibfnamefont {Y.-R.}\ \bibnamefont {Liu}}, \ and\ \bibinfo {author}
  {\bibfnamefont {S.-L.}\ \bibnamefont {Zhu}},\ }\href {\doibase
  10.1140/epjc/s10052-017-5297-4} {\bibfield  {journal} {\bibinfo  {journal}
  {The European Physical Journal C}\ }\textbf {\bibinfo {volume} {77}},\
  \bibinfo {pages} {709} (\bibinfo {year} {2017})}\BibitemShut {NoStop}%
\bibitem [{\citenamefont {Yang}\ \emph {et~al.}(2009)\citenamefont {Yang},
  \citenamefont {Deng}, \citenamefont {Ping},\ and\ \citenamefont
  {Goldman}}]{yangPhysRevD.80.114023}%
  \BibitemOpen
  \bibfield  {author} {\bibinfo {author} {\bibfnamefont {Y.}~\bibnamefont
  {Yang}}, \bibinfo {author} {\bibfnamefont {C.}~\bibnamefont {Deng}}, \bibinfo
  {author} {\bibfnamefont {J.}~\bibnamefont {Ping}}, \ and\ \bibinfo {author}
  {\bibfnamefont {T.}~\bibnamefont {Goldman}},\ }\href {\doibase
  10.1103/PhysRevD.80.114023} {\bibfield  {journal} {\bibinfo  {journal} {Phys.
  Rev. D}\ }\textbf {\bibinfo {volume} {80}},\ \bibinfo {pages} {114023}
  (\bibinfo {year} {2009})}\BibitemShut {NoStop}%
\bibitem [{\citenamefont {Feng}\ \emph {et~al.}(2013)\citenamefont {Feng},
  \citenamefont {Guo},\ and\ \citenamefont {Zou}}]{Feng:2013kea}%
  \BibitemOpen
  \bibfield  {author} {\bibinfo {author} {\bibfnamefont {G.~Q.}\ \bibnamefont
  {Feng}}, \bibinfo {author} {\bibfnamefont {X.~H.}\ \bibnamefont {Guo}}, \
  and\ \bibinfo {author} {\bibfnamefont {B.~S.}\ \bibnamefont {Zou}},\
  }\href@noop {} {\  (\bibinfo {year} {2013})},\ \Eprint
  {http://arxiv.org/abs/1309.7813} {arXiv:1309.7813 [hep-ph]} \BibitemShut
  {NoStop}%
\bibitem [{\citenamefont {Mehen}(2017)}]{MehenPhysRevD.96.094028}%
  \BibitemOpen
  \bibfield  {author} {\bibinfo {author} {\bibfnamefont {T.}~\bibnamefont
  {Mehen}},\ }\href {\doibase 10.1103/PhysRevD.96.094028} {\bibfield  {journal}
  {\bibinfo  {journal} {Phys. Rev. D}\ }\textbf {\bibinfo {volume} {96}},\
  \bibinfo {pages} {094028} (\bibinfo {year} {2017})}\BibitemShut {NoStop}%
\bibitem [{\citenamefont {Richard}\ \emph {et~al.}(2018)\citenamefont
  {Richard}, \citenamefont {Valcarce},\ and\ \citenamefont
  {Vijande}}]{Richard:2018yrm}%
  \BibitemOpen
  \bibfield  {author} {\bibinfo {author} {\bibfnamefont {J.-M.}\ \bibnamefont
  {Richard}}, \bibinfo {author} {\bibfnamefont {A.}~\bibnamefont {Valcarce}}, \
  and\ \bibinfo {author} {\bibfnamefont {J.}~\bibnamefont {Vijande}},\ }\href
  {\doibase 10.1103/PhysRevC.97.035211} {\bibfield  {journal} {\bibinfo
  {journal} {Phys. Rev.}\ }\textbf {\bibinfo {volume} {C97}},\ \bibinfo {pages}
  {035211} (\bibinfo {year} {2018})},\ \Eprint
  {http://arxiv.org/abs/1803.06155} {arXiv:1803.06155 [hep-ph]} \BibitemShut
  {NoStop}%
\bibitem [{\citenamefont {Cai}\ and\ \citenamefont
  {Cohen}(2019)}]{Cai:2019orb}%
  \BibitemOpen
  \bibfield  {author} {\bibinfo {author} {\bibfnamefont {Y.}~\bibnamefont
  {Cai}}\ and\ \bibinfo {author} {\bibfnamefont {T.}~\bibnamefont {Cohen}},\
  }\href@noop {} {\  (\bibinfo {year} {2019})},\ \Eprint
  {http://arxiv.org/abs/1901.05473} {arXiv:1901.05473 [hep-ph]} \BibitemShut
  {NoStop}%
\bibitem [{\citenamefont {Wang}\ \emph {et~al.}(2019)\citenamefont {Wang},
  \citenamefont {Liu},\ and\ \citenamefont {Liu}}]{Wang:2018atz}%
  \BibitemOpen
  \bibfield  {author} {\bibinfo {author} {\bibfnamefont {B.}~\bibnamefont
  {Wang}}, \bibinfo {author} {\bibfnamefont {Z.-W.}\ \bibnamefont {Liu}}, \
  and\ \bibinfo {author} {\bibfnamefont {X.}~\bibnamefont {Liu}},\ }\href
  {\doibase 10.1103/PhysRevD.99.036007} {\bibfield  {journal} {\bibinfo
  {journal} {Phys. Rev.}\ }\textbf {\bibinfo {volume} {D99}},\ \bibinfo {pages}
  {036007} (\bibinfo {year} {2019})},\ \Eprint
  {http://arxiv.org/abs/1812.04457} {arXiv:1812.04457 [hep-ph]} \BibitemShut
  {NoStop}%
\bibitem [{\citenamefont {Liu}\ \emph {et~al.}(2014)\citenamefont {Liu},
  \citenamefont {Li},\ and\ \citenamefont {Zhu}}]{LiuPhysRevD.89.074015}%
  \BibitemOpen
  \bibfield  {author} {\bibinfo {author} {\bibfnamefont {Z.-W.}\ \bibnamefont
  {Liu}}, \bibinfo {author} {\bibfnamefont {N.}~\bibnamefont {Li}}, \ and\
  \bibinfo {author} {\bibfnamefont {S.-L.}\ \bibnamefont {Zhu}},\ }\href
  {\doibase 10.1103/PhysRevD.89.074015} {\bibfield  {journal} {\bibinfo
  {journal} {Phys. Rev. D}\ }\textbf {\bibinfo {volume} {89}},\ \bibinfo
  {pages} {074015} (\bibinfo {year} {2014})}\BibitemShut {NoStop}%
\bibitem [{\citenamefont {Xu}\ \emph {et~al.}(2019)\citenamefont {Xu},
  \citenamefont {Wang}, \citenamefont {Liu},\ and\ \citenamefont
  {Liu}}]{Xu:2017tsr}%
  \BibitemOpen
  \bibfield  {author} {\bibinfo {author} {\bibfnamefont {H.}~\bibnamefont
  {Xu}}, \bibinfo {author} {\bibfnamefont {B.}~\bibnamefont {Wang}}, \bibinfo
  {author} {\bibfnamefont {Z.-W.}\ \bibnamefont {Liu}}, \ and\ \bibinfo
  {author} {\bibfnamefont {X.}~\bibnamefont {Liu}},\ }\href {\doibase
  10.1103/PhysRevD.99.014027} {\bibfield  {journal} {\bibinfo  {journal} {Phys.
  Rev.}\ }\textbf {\bibinfo {volume} {D99}},\ \bibinfo {pages} {014027}
  (\bibinfo {year} {2019})},\ \Eprint {http://arxiv.org/abs/1708.06918}
  {arXiv:1708.06918 [hep-ph]} \BibitemShut {NoStop}%
\bibitem [{\citenamefont {Li}\ \emph {et~al.}(2013)\citenamefont {Li},
  \citenamefont {Sun}, \citenamefont {Liu},\ and\ \citenamefont
  {Zhu}}]{PhysRevD.88.114008}%
  \BibitemOpen
  \bibfield  {author} {\bibinfo {author} {\bibfnamefont {N.}~\bibnamefont
  {Li}}, \bibinfo {author} {\bibfnamefont {Z.-F.}\ \bibnamefont {Sun}},
  \bibinfo {author} {\bibfnamefont {X.}~\bibnamefont {Liu}}, \ and\ \bibinfo
  {author} {\bibfnamefont {S.-L.}\ \bibnamefont {Zhu}},\ }\href {\doibase
  10.1103/PhysRevD.88.114008} {\bibfield  {journal} {\bibinfo  {journal} {Phys.
  Rev. D}\ }\textbf {\bibinfo {volume} {88}},\ \bibinfo {pages} {114008}
  (\bibinfo {year} {2013})}\BibitemShut {NoStop}%
\bibitem [{\citenamefont {Wang}\ and\ \citenamefont
  {Yan}(2018)}]{Wang:2017dtg}%
  \BibitemOpen
  \bibfield  {author} {\bibinfo {author} {\bibfnamefont {Z.-G.}\ \bibnamefont
  {Wang}}\ and\ \bibinfo {author} {\bibfnamefont {Z.-H.}\ \bibnamefont {Yan}},\
  }\href {\doibase 10.1140/epjc/s10052-017-5507-0} {\bibfield  {journal}
  {\bibinfo  {journal} {Eur. Phys. J.}\ }\textbf {\bibinfo {volume} {C78}},\
  \bibinfo {pages} {19} (\bibinfo {year} {2018})},\ \Eprint
  {http://arxiv.org/abs/1710.02810} {arXiv:1710.02810 [hep-ph]} \BibitemShut
  {NoStop}%
\bibitem [{\citenamefont {Agaev}\ \emph {et~al.}(2019)\citenamefont {Agaev},
  \citenamefont {Azizi}, \citenamefont {Barsbay},\ and\ \citenamefont
  {Sundu}}]{Agaev:2018khe}%
  \BibitemOpen
  \bibfield  {author} {\bibinfo {author} {\bibfnamefont {S.~S.}\ \bibnamefont
  {Agaev}}, \bibinfo {author} {\bibfnamefont {K.}~\bibnamefont {Azizi}},
  \bibinfo {author} {\bibfnamefont {B.}~\bibnamefont {Barsbay}}, \ and\
  \bibinfo {author} {\bibfnamefont {H.}~\bibnamefont {Sundu}},\ }\href
  {\doibase 10.1103/PhysRevD.99.033002} {\bibfield  {journal} {\bibinfo
  {journal} {Phys. Rev.}\ }\textbf {\bibinfo {volume} {D99}},\ \bibinfo {pages}
  {033002} (\bibinfo {year} {2019})},\ \Eprint
  {http://arxiv.org/abs/1809.07791} {arXiv:1809.07791 [hep-ph]} \BibitemShut
  {NoStop}%
\bibitem [{\citenamefont {Michael}\ and\ \citenamefont
  {Pennanen}(1999)}]{UKQCDPhysRevD.60.054012}%
  \BibitemOpen
  \bibfield  {author} {\bibinfo {author} {\bibfnamefont {C.}~\bibnamefont
  {Michael}}\ and\ \bibinfo {author} {\bibfnamefont {P.}~\bibnamefont
  {Pennanen}} (\bibinfo {collaboration} {UKQCD Collaboration}),\ }\href
  {\doibase 10.1103/PhysRevD.60.054012} {\bibfield  {journal} {\bibinfo
  {journal} {Phys. Rev. D}\ }\textbf {\bibinfo {volume} {60}},\ \bibinfo
  {pages} {054012} (\bibinfo {year} {1999})}\BibitemShut {NoStop}%
\bibitem [{\citenamefont {Detmold}\ \emph {et~al.}(2007)\citenamefont
  {Detmold}, \citenamefont {Orginos},\ and\ \citenamefont
  {Savage}}]{NPLPhysRevD.76.114503}%
  \BibitemOpen
  \bibfield  {author} {\bibinfo {author} {\bibfnamefont {W.}~\bibnamefont
  {Detmold}}, \bibinfo {author} {\bibfnamefont {K.}~\bibnamefont {Orginos}}, \
  and\ \bibinfo {author} {\bibfnamefont {M.~J.}\ \bibnamefont {Savage}}
  (\bibinfo {collaboration} {NPLQCD Collaboration}),\ }\href {\doibase
  10.1103/PhysRevD.76.114503} {\bibfield  {journal} {\bibinfo  {journal} {Phys.
  Rev. D}\ }\textbf {\bibinfo {volume} {76}},\ \bibinfo {pages} {114503}
  (\bibinfo {year} {2007})}\BibitemShut {NoStop}%
\bibitem [{\citenamefont {Bicudo}\ and\ \citenamefont
  {Wagner}(2013)}]{BicudoPhysRevD.87.114511}%
  \BibitemOpen
  \bibfield  {author} {\bibinfo {author} {\bibfnamefont {P.}~\bibnamefont
  {Bicudo}}\ and\ \bibinfo {author} {\bibfnamefont {M.}~\bibnamefont {Wagner}}
  (\bibinfo {collaboration} {European Twisted Mass Collaboration}),\ }\href
  {\doibase 10.1103/PhysRevD.87.114511} {\bibfield  {journal} {\bibinfo
  {journal} {Phys. Rev. D}\ }\textbf {\bibinfo {volume} {87}},\ \bibinfo
  {pages} {114511} (\bibinfo {year} {2013})}\BibitemShut {NoStop}%
\bibitem [{\citenamefont {Brown}\ and\ \citenamefont
  {Orginos}(2012)}]{BrownPhysRevD.86.114506}%
  \BibitemOpen
  \bibfield  {author} {\bibinfo {author} {\bibfnamefont {Z.~S.}\ \bibnamefont
  {Brown}}\ and\ \bibinfo {author} {\bibfnamefont {K.}~\bibnamefont
  {Orginos}},\ }\href {\doibase 10.1103/PhysRevD.86.114506} {\bibfield
  {journal} {\bibinfo  {journal} {Phys. Rev. D}\ }\textbf {\bibinfo {volume}
  {86}},\ \bibinfo {pages} {114506} (\bibinfo {year} {2012})}\BibitemShut
  {NoStop}%
\bibitem [{\citenamefont {Bicudo}\ \emph {et~al.}(2015)\citenamefont {Bicudo},
  \citenamefont {Cichy}, \citenamefont {Peters}, \citenamefont {Wagenbach},\
  and\ \citenamefont {Wagner}}]{bicudoPhysRevD.92.014507}%
  \BibitemOpen
  \bibfield  {author} {\bibinfo {author} {\bibfnamefont {P.}~\bibnamefont
  {Bicudo}}, \bibinfo {author} {\bibfnamefont {K.}~\bibnamefont {Cichy}},
  \bibinfo {author} {\bibfnamefont {A.}~\bibnamefont {Peters}}, \bibinfo
  {author} {\bibfnamefont {B.}~\bibnamefont {Wagenbach}}, \ and\ \bibinfo
  {author} {\bibfnamefont {M.}~\bibnamefont {Wagner}},\ }\href {\doibase
  10.1103/PhysRevD.92.014507} {\bibfield  {journal} {\bibinfo  {journal} {Phys.
  Rev. D}\ }\textbf {\bibinfo {volume} {92}},\ \bibinfo {pages} {014507}
  (\bibinfo {year} {2015})}\BibitemShut {NoStop}%
\bibitem [{\citenamefont {Bicudo}\ \emph {et~al.}(2016)\citenamefont {Bicudo},
  \citenamefont {Cichy}, \citenamefont {Peters},\ and\ \citenamefont
  {Wagner}}]{bicudoPhysRevD.93.034501}%
  \BibitemOpen
  \bibfield  {author} {\bibinfo {author} {\bibfnamefont {P.}~\bibnamefont
  {Bicudo}}, \bibinfo {author} {\bibfnamefont {K.}~\bibnamefont {Cichy}},
  \bibinfo {author} {\bibfnamefont {A.}~\bibnamefont {Peters}}, \ and\ \bibinfo
  {author} {\bibfnamefont {M.}~\bibnamefont {Wagner}},\ }\href {\doibase
  10.1103/PhysRevD.93.034501} {\bibfield  {journal} {\bibinfo  {journal} {Phys.
  Rev. D}\ }\textbf {\bibinfo {volume} {93}},\ \bibinfo {pages} {034501}
  (\bibinfo {year} {2016})}\BibitemShut {NoStop}%
\bibitem [{\citenamefont {Barnes}\ and\ \citenamefont
  {Swanson}(1992)}]{Barnes:1991em}%
  \BibitemOpen
  \bibfield  {author} {\bibinfo {author} {\bibfnamefont {T.}~\bibnamefont
  {Barnes}}\ and\ \bibinfo {author} {\bibfnamefont {E.~S.}\ \bibnamefont
  {Swanson}},\ }\href {\doibase 10.1103/PhysRevD.46.131} {\bibfield  {journal}
  {\bibinfo  {journal} {Phys. Rev.}\ }\textbf {\bibinfo {volume} {D 46}},\
  \bibinfo {pages} {131} (\bibinfo {year} {1992})}\BibitemShut {NoStop}%
\bibitem [{\citenamefont {Barnes}\ \emph
  {et~al.}(1999{\natexlab{b}})\citenamefont {Barnes}, \citenamefont {Black},
  \citenamefont {Dean},\ and\ \citenamefont {Swanson}}]{Barnes:1999hs}%
  \BibitemOpen
  \bibfield  {author} {\bibinfo {author} {\bibfnamefont {T.}~\bibnamefont
  {Barnes}}, \bibinfo {author} {\bibfnamefont {N.}~\bibnamefont {Black}},
  \bibinfo {author} {\bibfnamefont {D.~J.}\ \bibnamefont {Dean}}, \ and\
  \bibinfo {author} {\bibfnamefont {E.~S.}\ \bibnamefont {Swanson}},\ }\href
  {\doibase 10.1103/PhysRevC.60.045202} {\bibfield  {journal} {\bibinfo
  {journal} {Phys. Rev.}\ }\textbf {\bibinfo {volume} {C60}},\ \bibinfo {pages}
  {045202} (\bibinfo {year} {1999}{\natexlab{b}})},\ \Eprint
  {http://arxiv.org/abs/nucl-th/9902068} {arXiv:nucl-th/9902068 [nucl-th]}
  \BibitemShut {NoStop}%
\bibitem [{\citenamefont {Weinberg}(2005)}]{Weinberg:1995mt}%
  \BibitemOpen
  \bibfield  {author} {\bibinfo {author} {\bibfnamefont {S.}~\bibnamefont
  {Weinberg}},\ }\href@noop {} {\emph {\bibinfo {title} {{The Quantum theory of
  fields. Vol. 1: Foundations}}}}\ (\bibinfo  {publisher} {Cambridge University
  Press},\ \bibinfo {year} {2005})\BibitemShut {NoStop}%
\bibitem [{\citenamefont {Godfrey}\ and\ \citenamefont
  {Isgur}(1985)}]{Godfrey:1985xj}%
  \BibitemOpen
  \bibfield  {author} {\bibinfo {author} {\bibfnamefont {S.}~\bibnamefont
  {Godfrey}}\ and\ \bibinfo {author} {\bibfnamefont {N.}~\bibnamefont
  {Isgur}},\ }\href {\doibase 10.1103/PhysRevD.32.189} {\bibfield  {journal}
  {\bibinfo  {journal} {Phys. Rev.}\ }\textbf {\bibinfo {volume} {D 32}},\
  \bibinfo {pages} {189} (\bibinfo {year} {1985})}\BibitemShut {NoStop}%
\end{thebibliography}%

\end{document}